\documentclass[conference]{IEEEtran}
\IEEEoverridecommandlockouts
\usepackage{cite}
\usepackage[square,sort,comma,numbers]{natbib}
\usepackage{amsmath,amssymb,amsfonts}
\usepackage{algorithmic}
\usepackage{graphicx}
\usepackage{textcomp}
\usepackage{xcolor}
\usepackage{enumerate}
\usepackage{wrapfig}
\usepackage{bbm}
\usepackage{nicefrac} 
\usepackage{multirow}
\usepackage{multicol}
\usepackage{booktabs} 
\usepackage{adjustbox}

\def\BibTeX{{\rm B\kern-.05em{\sc i\kern-.025em b}\kern-.08em
    T\kern-.1667em\lower.7ex\hbox{E}\kern-.125emX}}
    
\DeclareMathOperator*{\argmax}{arg\,max}
\newcommand{\floor}[1]{\lfloor #1 \rfloor}

\begin{document}

\newcommand{\giorgi}[1]{\textcolor{blue}{[\textbf{Giorgi:} #1]}}
\newcommand{\ayu}[1]{\textcolor{cyan}{[\textbf{Ayu:} #1]}}
\newcommand{\nishan}[1]{\textcolor{olive}{[\textbf{Nishan:} #1]}}
\newcommand{\mohammad}[1]{\textcolor{red}{[\textbf{Mohammad:} #1]}}
         
\title{Two Is Better Than One: Dual Embeddings for Complementary Product Recommendations\\
}



\DeclareRobustCommand*{\IEEEauthorrefmark}[1]{%
  \raisebox{0pt}[0pt][0pt]{\textsuperscript{\footnotesize #1}}%
}

\author{%
 \IEEEauthorblockN{%
 Giorgi Kvernadze\IEEEauthorrefmark{1}\textsuperscript{\textsection},
 Putu Ayu G. Sudyanti\IEEEauthorrefmark{1}\textsuperscript{\textsection},
 Nishan Subedi\IEEEauthorrefmark{1},
 Mohammad Hajiaghayi\IEEEauthorrefmark{1}\IEEEauthorrefmark{,2}%
 }%
 \IEEEauthorblockA{\IEEEauthorrefmark{1} Overstock.com Inc., Salt Lake City, USA}%
 \IEEEauthorblockA{\IEEEauthorrefmark{2} University of Maryland, College Park, USA}
 giorgi.csv@gmail.com, pasudyan@gmail.com, nishansubedi@gmail.com, hajiaghayi@gmail.com
}%

\maketitle
\begingroup\renewcommand\thefootnote{\textsection}
\footnotetext{Co-first authors.}

\thispagestyle{plain}
\pagestyle{plain}

\begin{abstract}
    Embedding based product recommendations have gained popularity in recent years due to its ability to easily integrate to large-scale systems and allowing nearest neighbor searches in real-time. The bulk of studies in this area has predominantly been focused on similar item recommendations. Research on complementary item recommendations, on the other hand, still remains considerably under-explored. 
    We define \textit{similar} items as items that are interchangeable in terms of their utility and \textit{complementary} items as items that serve different purposes, yet are compatible when used with one another. 
    In this paper, we apply a novel approach to finding complementary items by leveraging dual embedding representations for products. We demonstrate that the notion of relatedness discovered in NLP for \textit{skip-gram negative sampling (SGNS)} models translates effectively to the concept of complementarity when training item representations using co-purchase data. Since sparsity of purchase data is a major challenge in real-world scenarios, we further augment the model using synthetic samples to extend \textit{coverage}. This allows the model to provide complementary recommendations for items that do not share co-purchase data by leveraging other abundantly available data modalities such as images, text, clicks etc. We establish the effectiveness of our approach in improving both coverage and quality of recommendations on real world data for a major online retail company. We further show the importance of task specific hyperparameter tuning in training SGNS. Our model is effective yet simple to implement, making it a great candidate for generating complementary item recommendations at any e-commerce website. 
\end{abstract}



\section{Introduction}
\label{sec:intro}

Product recommendations have become one of the most important tools in e-commerce for product discovery, and have proven to drive significantly increased sales in various settings. A study conducted by Barilliance Research estimated that up to 31\% of total website revenue and, on average, 12\% of sales were associated with product recommendations. For website users that clicked on recommended items, they were five times more likely to make a purchase than those who did not. Additionally, recommendation systems provide a better overall customer experience, which leads to better retention rates. In general, we encounter two different types of recommendations: similar and complementary recommendations. \textit{Similar} product recommendation provides relevant items that have identical functionality (e.g. brown sofas and black sofas). Due to its interchangeability, customers often purchase one over the other, but would rarely purchase both. The goal of showing this type of recommendation is then to convince users to eventually make a purchase by providing a diverse set of related items to their original intended product of interest. On the other hand, \textit{complementary} product recommendation aims to suggest products that serve a different purpose but are compatible when used in conjunction (e.g. mattresses and mattress pads). This type of recommendation has been shown to contribute significantly to increasing conversion as well as average order value in e-commerce websites. Both types can be customized at the user level, creating more personalized recommendations for each user based on their past interactions with the website (e.g. purchases, browsing, add-to-carts, etc.).

We focus our attention on the generation of complementary recommendations. Extensive research has been done in the area of similar product recommendation from the classic collaborative filtering methods \citep{ekstrand2011collaborative, schafer2007collaborative, koren2009collaborative}, to the widely implemented matrix factorization model \citep{koren2009matrix, mnih2007probabilistic, xue2017deep}, and most recently, to the use of vector representation learned from different data modalities through various neural network architecture \citep{mcauley2015image, grbovic2015commerce, angelovska2021siamese}. 
While there is substantial research in the area of similarity recommendation, limited work can be found for complementary recommendations. As we discuss below, the problem of finding products that are complementary to each other is an inherently more difficult task than finding items that are interchangeable in nature. Unlike similar product recommendation, complementary products adheres to additional rules such as asymmetry where the direction of the recommendations is typically relevant in its evaluation. One example would be phones and screen protectors. Customers who purchase a phone would likely be interested in buying screen protectors, however, they would rarely purchase a phone when looking to buy screen protectors. Additionally, the transitive property between the recommended products does not always hold in complementary relationship. A coffee table would be a good complementary item for a sofa, a coaster is a good complementary item for a coffee table, but a coaster would be less relevant to a sofa. 

The earliest work in discovering complementary relationships centered upon mining historical purchases to find patterns of products that are frequently bought together while accounting for other additional measures \citep{agrawal1993mining}. Although simple, effective, and interpretable, this method suffers from inefficiencies in large data sets as well as the cold-start problem for items that were never purchased. Another approach is through the use of a supervised learning methods. Manually labeled sets of products can be used as positive examples that are feed into a classifier to build models that learn to identify patterns of complementarity between pairs of products \citep{angelovska2021siamese}. One advantage of this approach is that it allows learning from different sources of data. However, human labeled data is expensive to obtain and thus sparsely available, leading to limitations in training certain models. For this reason, most studies substitute the use of human labeled data with historical customers' behavior data such as co-purchases, views, and co-clicks \citep{mcauley2015image}. These labels come with their own noise and biases, but as we show in this paper, a careful pre-processing of the data can alleviate some of these issues. 

A related line of research on complementary recommendations involves the bundling of a set of products, where business components such as price and discount savings as well as other constraints are contributing factors in the optimization of the models \citep{zhu2014bundle, bai2019personalized}. These studies are disparate from our goal of solely focusing on finding complementary relationships between goods without constraints on factors such as discounts. 

We concentrate on the latent representation approach to producing complementary recommendations.
Representation learning has been widely applied to predict similarity between items. This framework uses low-dimensional vectors to summarize information about an entity. These vectors can be generated from different data modalities such as images \citep{mcauley2015image}, text \citep{mcauley2015inferring}, or any other relevant signals.
One of the most commonly used models for learning vector representations for products is the {\it Prod2Vec} model \citep{grbovic2015commerce, barkan2016item2vec}, it has gained attention as a method for product recommendations due to its simplicity, efficiency, and effectiveness. The main idea behind this model is to use the {\it Word2Vec} algorithm \cite{mikolov2013distributed, mikolov2013efficient} on sequences of products. The {\it Word2Vec} model uses a shallow neural network architecture to learn low-dimensional vector representation of words from sequences of text, such that semantically similar words are placed close to each other in the embedding space. In the {\it Prod2Vec} model, the sequences of words translate to sequences of user actions on products, e.g., clicks, add-to-carts, purchases etc. A model trained on such data will learn to represent products in a lower-dimensional space that captures certain semantic relationships between products. For example, if two products often appear in similar contexts, the vectors representing these products will lie close to each other in the embedding space. This effectively reduces the problem of producing similar items recommendations to a nearest neighbor search using simple distance functions like inner product or cosine similarity. The existence of numerous fast approximate nearest neighbor search algorithms \cite{wang2021comprehensive, FuNSG17, 2205.03763} makes this method particularly attractive, since it can scale to large datasets, even in settings where inference is done in real-time.


While finding substitute items is a straightforward procedure using the {\it Prod2Vec} model, it is not as clear whether we can produce complementary recommendations in the same manner. If we train the model on sequences of products that reflect complementary relationships, commonly found in co-purchase data, the assumption is that these relationships will be uncovered in the resulting embedding space. However, we  show in this paper that the process is more involved. The {\it Word2Vec} architecture produces two weight matrices: the input and output matrix. In practice, only one of these matrices is being used (input) as the vector representation of the entities. Several studies \cite{asr2018querying, mitra2016dual} in NLP have found that the dot product between two word embeddings from the two different matrices provides an indicator for \textit{relatedness}, whereas that of the same matrix provides a measure of similarity. In this paper, we explore this relationship as we translate relatedness in NLP to complementarity in recommendation systems. Our contribution is threefold:

\begin{enumerate}
    \item We describe a simple at-inference-time adaptation of SGNS that transforms the model into a highly efficient and effective complementary product candidate retrieval method.
    \item We introduce a practical data augmentation method for overcoming the challenge of \textit{coverage}, often encountered when dealing with purchase data. This allows the model to provide complementary candidate sets for items that do not share co-purchase data by utilizing similarity measurements derived from other types of abundantly available data sources, such as clicks or product metadata. Our results show that our method not only substantially expands the coverage, but also improves the relevancy of the candidates.
    \item We show the importance of tuning the hyperparameters of SGNS as opposed to using the default hyperparameters passed down from the original implementation \cite{mikolov2013distributed}. We show an upwards of 300\% improvement on relevant metrics when tuning the hyperparameters on the appropriate downstream task. 
\end{enumerate}

In the next section, we will discuss related research in the area. In Section \ref{sec:method} we present our proposed method followed by experimentation and results in Section \ref{sec:experiments}. Lastly, we will summarize and conclude our finding in the final section (Section \ref{sec:discussions}) of this paper.

\section{Related Literature
\label{sec:litreview}}

Since the introduction of the {\it Word2Vec} model \cite{mikolov2013distributed, mikolov2013efficient}, a variety of models have been proposed for its adaptation to recommender systems. The {\it Prod2Vec} and {\it bagged-Prod2Vec} \citep{grbovic2015commerce} were among the first models proposing this adoption, where email receipts of purchases were used to create a sequence of products for learning product vector representation for recommendation. A variation of that which involves textual information of the products were introduced in \citep{vasile2016meta}. Graph-based approaches \citep{grover2016node2vec, wang2018billion} were proposed by incorporating the network structure between different entities (products, users, taxonomy, etc.) and learning the latent space from random walks sampled from the distribution of this network. 


While the majority of the literature focuses on generating similar recommendations under the {\it Prod2Vec} framework, very little has been done on exploring how the model performs in generating complimentary recommendations. One study that has attempted to do this in the grocery shopping domain is the {\it Triple2Vec} model \citep{wan2018representing}. The authors proposed creating triplets of {\it (item, item, user)} from the user shopping basket to recover the complementary, compatibility, and loyalty relationship of products and users. However, the paper failed to explore the relationship between the two resulting matrices from the model and resort to the conventional method of using one matrix as embeddings while discarding the other. Additionally, it does not mention ways of tackling the cold-start problem for items that were never purchased. 

{\it BB2vec} \citep{trofimov2018inferring} proposed the use of baskets and browsing session data to create the product embeddings. A multi-task learning layer were used on top of the learned representation and assumed some representations are shared between learning tasks, which represent different types of data sources (browsing and baskets). While the author touched upon the use of both resulting matrices from the model, there were limited explanation of why this mechanism works. Further, multi-task learning can be difficult to train, especially on browsing data that are inherently noisy. 

Other studies such as \cite{chen2020studying} demonstrated the use of both types of matrices to compute complementarity between two products. Again, the authors failed to provide explanation on why this method works in finding the complementary relationship. We seek to fill this void in the literature by providing a thorough explanation backed by experiments conducted on two real world data sets from major online retail companies.

\section{Related Models}
\label{ref:prod2vec}


In the following section we first briefly introduce the Skip-gram model made popular by {\it Word2Vec}, we then describe how we adapt the model to our specific problem. Finally, we show how the model can be used to infer complementary relationships. 

\subsection{Prod2Vec: Product Representation Learning}
The Skip-gram model was first introduced in NLP as a technique to learn vector representations of words, but has since been adopted into various other domains \cite{grbovic2015commerce, du2019gene2vec, choi2016multi, vasile2016meta}. In e-commerce, it can be used to learn low dimensional product representations by training the model on a sequence of user actions e.g. clicks, add-to-cart, purchases, add-to-wish-list etc. The training objective of Skip-gram is to learn to represent items in such a way that it becomes a good predictor of its surrounding items in a sequence. More formally, given a collection of sequences $S$, where each sequence is composed of products $p$ in the vocabulary $P$, the objective is to maximize the log-likelihood:
\begin{equation}
\argmax_{\mathbf{W}_{in} , \mathbf{W}_{out}} \sum_{s \in S} \sum_{p_{i} \in s} \sum_{-w \leq j \leq w, w \neq 0} \log{\mathbb{P}{(p_{i + j} \mid p_{i}})}
\end{equation}
Where $\mathbf{W}_{in}$ and $\mathbf{W}_{out}$ are the input and output embeddings respectively and $w$ is a hyperparameter defining the size of the context window. The conditional probability of observing a context product given a target product is computed as:
\begin{equation}
\mathbb{P}{(p_{c} \mid p_{t}}) = \frac{\exp{(\mathbf{v}_{p_{t}}^T \cdot \mathbf{v}^{'}_{p_{c}})}}{\sum_{m = 1}^{|P|} \exp{(\mathbf{v}_{p_{t}}^T \cdot \mathbf{v}^{'}_{m})}}
\end{equation}

where $\mathbf{v}_{p}, \mathbf{v}^{'}_{p} \in \mathbb{R}^{d}$ are the $d$ dimensional input and output vectors for the product $p$. Computing the normalization factor of the conditional probability can be expensive with large vocabularies. A popular approach is to instead optimize a different objective function that approximates the softmax introduced by Mikolov et al. \cite{mikolov2013distributed}. The resulting model for $\mathbb{P}{(p_{c} \mid p_{t}})$ is often referred to as Skip-gram with negative sampling or SGNS for short and is defined as follows:
\begin{equation}
\log{\sigma{(\mathbf{v}_{p_{t}}^T \cdot \mathbf{v}^{'}_{p_{c}}})} + \sum_{i = 1}^{k} \mathbb{E}_{p_{i} \sim N(p)} \log{\sigma{(-\mathbf{v}_{p_{t}}^T \cdot \mathbf{v}^{'}_{p_i}})}
\end{equation}

In order to simplify both the data pre-processing and training procedure, we process the purchase sequences into purchase pairs by taking all pairwise combinations in each session. That is, for a given session $s = \{p_1, p_2, p_3, ..., p_n\}$ we take the Cartesian product with itself $s \times s = \{(p_i, p_j) \mid p_i \in s, p_j \in s, p_i \neq p_j \}$. By doing this with each session, we get a new pairwise co-purchase data $D$ and the objective function defined in (1) becomes:

\begin{equation}
\argmax_{\mathbf{W}_{in} , \mathbf{W}_{out}} \sum_{(p_{i}, p_{j}) \in \mathcal{D}} \log{\mathbb{P}{(p_{i} \mid p_{j}})} + \log{\mathbb{P}{(p_{j} \mid p_{i}})}
\end{equation}

Having the data defined in this form allows us more flexibility with filtering based on heuristics, e.g. removing pairs that have identical taxonomy hierarchies or removing pairs that do not have a strong enough connection in terms of co-occurrences (more on this in a later chapter about data preprocessing in Section  \ref{sec:data-prep}). It also allows us to discard the hyperparameter $w$, since context window size is no longer relevant. 

\subsection{Dual Embedding Model for Relatedness}
The feed forward neural network architecture of the {\it Word2Vec} model consist of two different weight matrices: the input (word) and output (context) matrix. Once training is completed and the model is optimized, the common practice is to discard the output matrix and use the input matrix as the final product embeddings. Subsequent similarity search tasks between products are then performed through the vector representation of each product in this embedding space. Even though this works effectively in practice, little is known on the additional information uncovered on the distribution of joint behavior of these two matrices. Simple combination of the two vectors such as concatenation, addition \citep{pennington2014glove}, or averages \cite{levy2015improving} have been investigated, more thorough studies are still needed to prove its significance in boosting performance over using the standalone input matrix. On the other hand, multiple studies in NLP have found evidence to support that using the input vectors as word embeddings predicts better similarity, while using a combination of the two vectors resulted in improved relatedness between words. An example of pairs with high similarity score would be "sea" and "ocean", and pairs with high relatedness score would be "ocean" and "coast". 

The {\it Dual Embedding Space Model (DESM)} \citep{mitra2016dual} explores this relationship for search and document retrieval task. The authors investigated the relatedness aspect between a query word and all the terms in the document. They found that by using both the input and the output representations jointly, they were able to better rank the \textit{aboutness} aspect of a document with respect to a query term. The authors discovered that the embeddings of words in the input vector space tend to be closer to the output vector representation of words that would co-occur together. This mean that the cosine similarity between words amongst the input-input and output-output vector space are higher for words that are functionally similar, whereas in cross-vector relationship, input to output vector space, the similarity are higher for words that appear together often in the training sample. One of their example was for the word "Yale". The neighborhood of the input vector of "Yale" in the input vector space corresponds to words like "Harvard", "NYU", "Tulane", and "Tufts" which are words that would be found in a similar context to the query word "Yale". A comparable pattern was found for the output vector in the neighborhood of the output vector space. In contrast, if we look at the neighborhood of the input vector of the word "Yale" in the output vector space, they find words like "faculty", "alumni", and "graduate" which are related to the query word and would appear together in a sentence but would not be used in the same context. 

Another relevant study in support of the idea of using both the input and output embeddings to measure relatedness between words was done by Asr et al. \cite{asr2018querying}. The authors used two distinct word data sets, where one was designed for participants to exclusively measure the degree of similarity between pairs of words (e.g. "sea" and "ocean") and the other for relatedness (e.g. "clothes" and "closet").  More formally, let $\mathbf{v}_{in}^w$ and $\mathbf{v}_{out}^w$ be the input and output embeddings for any word $w \in V$ where $V$ is the vocabulary of all words and let $(w_i, w_j)$ be pairs of words in the data sets for $w_i, w_j \in V$. They found that the cosine similarity between $\mathbf{v}_{in}^{w_i}$ and $\mathbf{v}_{in}^{w_j}$ across all pairs have the highest Spearman correlation to the human similarity judgements data set. On the other hand, the cosine similarity between $\mathbf{v}_{in}^{w_i}$ and $\mathbf{v}_{out}^{w_j}$ across all pairs have the highest correlation to the relatedness scores of the second data set. In particular, they discovered that the cosine similarity between $\mathbf{v}_{in}^{w_i}$ and $\mathbf{v}_{out}^{w_j}$ performed better than that of $\mathbf{v}_{out}^{w_i}$ and, $\mathbf{v}_{in}^{w_j}$ which suggest that the likelihood of seeing $w_j$ in the context of $w_i$ is higher than the likelihood of seeing $w_i$ in the context of $w_j$. This indicates and signifies that the model is able to recognize the notion of forward relatedness and asymmetry of the pairs. 

In the following section, we describe how we can translate the concept of dual embeddings into the recommendation systems domain, specifically in the generation of complementary product recommendations. 

{\renewcommand{\arraystretch}{1.25}
\begin{table*}[!ht]
\caption{Comparison between the dual embeddings model (\textbf{IN-OUT}) against baseline models.}
\centering
    \begin{tabular}{l c c c c c c c c}
    \hline
         \multirow{2}{*}{\bf Model} & \multicolumn{4}{c}{\bf Overstock} & \multicolumn{4}{c}{\bf Instacart} \\\cline{2-9}
         & Precision@20 & Recall@20 & Precision@50 & Recall@50 & Precision@20 & Recall@20 & Precision@50 & Recall@50 \\
         \hline
         Co-Purchases & 0.0348 & 0.2745 & 0.0153 & 0.2800 & 0.0409 & 0.0514 & 0.0276 & 0.0792 \\
         Top-Sellers & 0.0149 & 0.1029 & 0.0073 & 0.1141 & 0.0305 & 0.0698 & 0.0174 & 0.1156 \\
         IN-IN & 0.0142 & 0.1244 & 0.0073 & 0.1514 & 0.0138 & 0.0136 & 0.0099 & 0.0237 \\
         OUT-OUT & 0.0187 & 0.1364 & 0.0104 & 0.1694 & 0.0163 & 0.0117 & 0.0134 & 0.0199 \\
         IN-OUT & {\bf 0.0391} & {\bf 0.3229} & {\bf 0.0187} & {\bf 0.3523} & {\bf 0.0437} & {\bf 0.0702} & {\bf 0.0322} & {\bf 0.1226}\\
         \hline
    \end{tabular}
    \label{tab:baseline_res}
\end{table*}}

\begin{figure}
    \centering
    \includegraphics[scale=0.28]{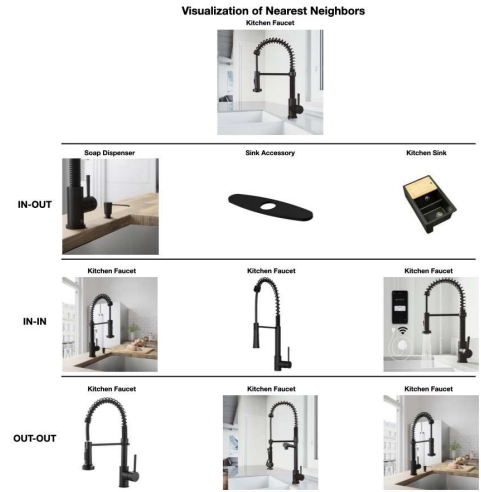}
    \caption{Example recommendations produced by different dot products of the input and output vectors. Given kitchen faucet as a target item, the {\bf IN-OUT} variation provides recommendations that are complementary to the target (e.g. soap dispensers, sink accessories, kitchen sinks). In contrast, both the {\bf IN-IN} and {\bf OUT-OUT} variation resulted in similar recommendations (other kitchen faucets).}
    \label{fig:recs_compare}
\end{figure}

\section{Proposed Method}
\label{sec:method}

In this section, we describe our proposed method for complementary product recommendations, as well as our data augmentation approach to expand product coverage to cold-start items. 

\subsection{Dual Embeddings for Complementary Product Retrieval (DE)}


As previously mentioned, one of the key details that is often looked over in SGNS models is the fact that every item has two representations. For every item $p$, the model produces two vectors $\mathbf{v}_{p}$,  $\mathbf{v}^{'}_{p}$ corresponding to that item, where $\mathbf{v}_{p}$ and $\mathbf{v}^{'}_{p}$ are the vectors contained in the input embedding matrix $\mathbf{W}_{in}$ and output embedding matrix $\mathbf{W}_{out}$, respectively. Once the model is trained, practitioners usually  discard one of the vector representations and use the other for inference, a popular choice is to keep the vectors in the input embedding matrix $\mathbf{W}_{in}$. In order to retrieve candidates for a target item $p$, we then find the items that correspond to the highest cosine similarity score:
\begin{equation}
\argmax_{\mathbf{v} \in \mathbf{W}_{in}} cosine(\mathbf{v}_{p}, \mathbf{v})
\end{equation}
While this is valid for problems where the goal is to capture similarity between two items, it is not as effective when we require co-occurrence relationships such as complementarity. This is because the objective of the SGNS model is to maximize the likelihood of predicting the surrounding items given a target item, where the surrounding items are represented using the outputs vectors and the target item is represented using the input vector. This causes the model to learn to represent items within each of the embedding spaces ($\mathbf{W}_{in}$ and $\mathbf{W}_{out}$) in a way that makes them a good predictor for their neighbors, meaning that items that share neighbors or contexts will be represented similarly, implying substitutionary relationships. On the other hand, the model also learns to place items that often co-occur together near each other across the embedding spaces, since this is what is directly being optimized at training time. Hence, we should be expecting mostly similar relationships being captured within each of the embedding spaces but co-occurrence based relationships (e.g. complementarity) across the two embedding spaces. This is consistent with what we see in the NLP setting (Section \ref{ref:prod2vec}) where the use of both vectors to compute the cosine similarity resulted in a better indication of relatedness between words, whereas the use of a single vector is best for measuring similarity.


We can translate the concept of relatedness between words to the concept of complementarity between products. In particular, we can best illustrate this relationship using co-purchase pairs. Given two pairs of co-purchased items, $r_1$ and $r_2$ where $r_1 = (queen\:mattress, bed\:sheet)$ and $r_2 = (twin\:mattress, bed\:sheet)$. The resulting input matrix of the model will be able to capture that $queen\:mattress$ is similar to, $twin\:mattress$ since they both appear in a similar "context" to $bed\:sheet$. In the same way, the resulting matrices will also be able to recognize that $bed\:sheet$ is \textit{related} or \textit{complementary} to $queen\:mattress$ and $twin\:mattress$ as they co-occur. 

For the rest of the paper, we introduce the following notation to denote the different sets of recommendations generated using the input and output vectors of a given product $p$: 
\begin{equation*}
\textbf{IN-OUT}: \argmax_{\mathbf{v}^{'} \in \mathbf{W}_{out}} cosine(\mathbf{v}_{p}, \mathbf{v}^{'})\label{eq:in-out}
\end{equation*}
\begin{equation*}
    \textbf{IN-IN}: \argmax_{\mathbf{v} \in \mathbf{W}_{in}} cosine(\mathbf{v}_{p}, \mathbf{v})
\end{equation*}
\begin{equation*}
    \textbf{OUT-OUT}: \argmax_{\mathbf{v^{'}} \in \mathbf{W}_{out}} cosine(\mathbf{v}_{p}^{'}, \mathbf{v^{'}})
\end{equation*}
\begin{equation*}
    \textbf{OUT-IN}: \argmax_{\mathbf{v} \in \mathbf{W}_{in}} cosine(\mathbf{v}_{p}^{'}, \mathbf{v})
\end{equation*}

To capture complementary relationships, we will be relying on inference using \textbf{IN-OUT}. In our experiments, we do not consider the \textbf{OUT-IN} variant because we are concerned with the forward relationship between pairs of products (e.g. $bed\:sheet$ as a complementary item to $queen\:mattress$ but not the other way around). This is consistent with the study conducted by Asr et al. \cite{asr2018querying} on forward relatedness.


\subsection{Data Augmentation (DA)}

One of the main issues of working with purchase data is sparsity. Purchase data on products isn't as widely available as other types of user generated signals like clicks or add-to-carts. Furthermore, content data such as product title, description or images are usually available for all products. The question we want to investigate is, how can we use these other product signals to augment the data on purchases? More specifically, we want a method that can use existing co-purchased product pairs to synthesize new product pairs that are not yet observed in the training set. Inspired by related work in NLP on data augmentation for classification tasks \cite{wang-yang-2015-thats, wei-zou-2019-eda}, we introduce a simple yet effective method for data augmentation for complementary recommendations.

Given a similarity function that can provide a real-valued score for any given pair of products, we use the function to infer possible new pairwise relationships not yet observed in the training data by creating pairs that are similar to existing ones. For example, let $(p_{a}, p_{b})$ be a real co-purchased pair and $p_{a'}$ and $p_{b'}$ be products that have high similarity to $p_{a}$ and $p_{b}$ respectively, then we create new pairs $(p_{a'}, p_{b})$, $(p_{a}, p_{b'})$ and add it to the dataset. We can similarly create a new pair by replacing both of the original products by the similar items of each to get $(p_{a'}, p_{b'})$. This idea relies on a simple fact that if two items are similar then they probably complement the same item, e.g., two very similar lamps should both complement the same couch. 

One key question is what co-purchase count do we assign to the synthetic pairs? We think a reasonable choice would be to rely on the original co-purchase count $c_{ab}$ of $(p_{a}, p_{b})$ and adjust it according to the strength of similarity to the new product. For example, for the new pair, $(p_{a'}, p_{b})$ we assign co-purchase count $c_{a'b} = \floor{c_{ab} \cdot s(a, a')}$ where $s(., .) \in [0, 1]$ is a similarity function. The justification for the similarity score multiplier is that highly similar products should have comparable co-purchase counts, whereas items that are not as similar should be penalized. 

For completeness, we define the two operations to synthesize new pairs below. Note that we introduce an additional hyperparameter $\gamma \in (0, 1)$ that is used to tamper down some effects the synthetic pairs will have on the entire dataset, since if synthetic pairs are added with too high of co-purchase counts this can dominate the entire dataset and could potentially have a negative impact on performance.


\begin{enumerate}
    \item \textbf{Replace Single}: We replace one of the members of the real co-purchase pair by its most similar item, e.g., we can create a synthetic pair $(p_{a'}, p_{b})$ with a co-purchase count set to $c_{a'b} = \floor{\gamma \cdot c_{ab} \cdot s(a, a')}$. We can similarly create a pair $(p_{a}, p_{b'})$.
    \item \textbf{Replace Both}: We replace both of the products in the pair with their respective similar items to get $(p_{a'}, p_{b'})$ with a co-purchase count of $c_{a'b'} = \floor{\gamma \cdot c_{ab} \cdot \frac{s(a, a') + s(b, b')}{2}}$.
\end{enumerate}


Although the method we described creates synthetic pairs using only the most similar one item to the existing one, in practice this can be replaced by $k$ most similar items which can allow for an upwards of $k^2 + 2k$ synthetic pairs to be created using only a single real pair. In practice, we would also recommend introducing a parameter that defines the minimum threshold $\theta$ on the similarity scores, as this can allow for better control on the quality of the added synthetic pairs.

\subsection{Inference Augmentation (IA)}

We can apply a similar idea at inference time as well. That is, for a given target product that is not represented in our training set, meaning that we are not able to generate recommendations for it, we can first look up the most similar item that does appear in the training set and use it to infer complementary products on the original target item. Let $p_{a'}$ be a target product that does not exist in our training set, we first retrieve a product $p_{a}$ from our training set such that $p_{a} = \argmax_{p \in P} s(a', p)$ and then we proceed to recommend items using \eqref{eq:in-out}. We refer to this method as inference augmentation or IA for short.


\section{Experiments and Results}
\label{sec:experiments}

{\renewcommand{\arraystretch}{1.25}
\begin{table*}[!ht]
\caption{The effects of data and inference augmentation on different subset of the validation set}
\centering
    \begin{tabular}{l c c c c c c}
    \hline
         \multirow{2}{*}{\bf Model} & \multicolumn{2}{c}{\bf In-Coverage} & \multicolumn{2}{c}{\bf Out-of-Coverage} & \multicolumn{2}{c}{\bf Combined} \\ \cline{2-7}
         & Precision@20 & Recall@20 & Precision@20 & Recall@20 & Precision@20 & Recall@20 \\
         \hline
         Random & 0.0093 & 0.0383 & 0.0006 & 0.0052 & 0.0059 & 0.0262 \\
         Top-Sellers & 0.0216 & 0.1419 & 0.0017 & 0.0253 & 0.0149 & 0.1029 \\
         IN-OUT &  0.0584 & 0.4847 & 0.0000 & 0.0000 & 0.0388 & 0.3224 \\
         IN-OUT+DA & \textbf{0.0589} & \textbf{0.4912} & 0.0012 &  0.0163 & 0.0396   & 0.3322 \\
         IN-OUT+IA & N/A & N/A & 0.0018 & 0.0251  & 0.0368 & 0.3232 \\
         IN-OUT+DA+IA & N/A & N/A & \textbf{0.0038} & \textbf{0.0269} & \textbf{0.0398} & \textbf{0.3345}\\
         \hline
    \end{tabular}
\label{table:da-eval}
\end{table*}
}

We compare the performance of our model against the following baselines:
\begin{itemize}
    \item {\bf Top-Sellers}  Recommend most popular (the highest selling) items for every target item. This heuristic can be effective for items that are complementary to many other items, e.g., eggs and milk in grocery shopping data. 
    \item {\bf Co-Purchases} For a given target item, recommend most frequently co-purchased items (in descending order). This model is typically used as a baseline model, as it generally produces high relevancy. 
    \item {\bf IN-IN \& OUT-OUT} Recommend nearest neighbors of the target item based on either the input or output vectors.
\end{itemize}

\subsection{Datasets}
We evaluate our models on real world data from two major online retail companies: 
\begin{itemize}
    \item {\bf Overstock.com}: Proprietary data from major online home-goods retail company. This dataset consists of two years worth of transactions made through the Overstock.com website. All orders in the most recent month are used as held out test set and the rest are used for training. It contains over 18 million users across 24 million sessions with around 1.6 million distinct product IDs. The product IDs spans beyond 3 thousands different sub-categories and 168 departments. 
    \item {\bf Instacart.com} \cite{instacart}: Publicly available data from a major grocery delivery company. This data contains over 3 million orders across 21 departments, 134 different aisles from more than 200 thousands users. The dates of the transactions were not included, however, the add-to-cart order were part of the additional information. We used the training set that were made available and pre-process the data accordingly.
\end{itemize}

\subsection{Implementation Details}

To train the SGNS model we use \textit{fastText++} \cite{fastText++} which is an extended version of the popular library \textit{fastText} \cite{bojanowski2016enriching} developed by Facebook AI Research lab. We disable the subword embedding option, as we do not have text as input. All the hyperparameters of the model are optimized using \textit{Ray Tune} \cite{liaw2018tune} on a cluster of e2-highcpu-32 machines on Google Cloud Platform, more details on hyperparameter tuning are provided in a later section. At inference time, for fast approximate nearest neighbor searches, we use Hierarchical Navigable Small World (HNSW) \cite{hnsw2020} graphs as implemented in \textit{Faiss} \cite{johnson2019billion}. 

\subsection{Data Pre-processing}
\label{sec:data-prep}

We remove overly active users (higher number of unique purchases than 99.9\% population), as these users are likely associated with drop-shipping. We similarly remove purchase sessions that have a high number of unique purchases (again above 99.9\%) since a high number of purchases in a session potentially provides a weaker signal on the relationship between pairs of items in that session. We create a pairwise product dataset from the purchase sessions by taking all unique pairs in a given purchase session. We remove any product pair that does not have at least 3 co-purchases. We measure Pointwise Mutual Information (PMI) \cite{church-hanks-1990-word}, as defined below, to further identify pairs that provide weaker signal on complementarity by removing all pairs with negative PMI. 

\begin{equation}
    PMI(p_i, p_j) = \log \left(\frac{n_{ij}/T}{(n_i/T)(n_j/T)}\right)
\end{equation}
where $n_{i}$ is the number of times the product $p_i$ was purchased, $n_{ij}$ is the number of times the pair ($p_i, p_j$) was co-purchased, and $T$ is the total number of sessions. 

We also discard product pairs that have identical taxonomies, since these usually coincide with similar products rather than complementary. There are some exceptions to this rule (e.g. throw pillows) which we account for by creating a taxonomy exception list curated from highly purchased pairs from the same taxonomy. Similar to the product pairwise data, we also create a taxonomy dataset aggregated in pairwise manner. We compute PMI between the pairs of taxonomies, which allows us to use it for another round of filtering at the product level, i.e. if a product pairs' taxonomy PMI is lower than a minimum threshold, we discard that pair. The aforementioned processing technique is especially important for filtering synthetic pairs at the data augmentation stage because those pairs do not have actual purchases associated with it, but we can still rely on taxonomy level scores for discarding noisy pairs.

\subsection{Dual Embeddings Evaluation}

We evaluate our models on a held out test set processed in the same way as our training sets. Pairs of products were created from the purchase data, then for each query product $q \in Q$, where $Q$ is the sets of all query products, we ranked the co-purchase product IDs based on the number of times they co-occur. Let $L_q$ be the list of ranked co-purchase products for the item $q$ deduced from the test set, and $R_q^K$ be the top $K$ predicted list of complementary product recommendations for the item $q$ produced by the model. As our goal for the model is to retrieve candidate sets, we focus on two main metrics that best align with this goal: {\it Precision@K}  and {\it Recall@K}. We define these metrics as follows:
\begin{equation*}
    Precision_q@K = \frac{|L_q \cap R_q^K|}{K}, \quad Recall_q@K = \frac{|L_q \cap R_q^K|}{|L_q|}
\end{equation*}
We report the mean of both metrics $\forall q \in Q$ in our test set and choose $K \in \{20, 50\}$. We see from Table \ref{tab:baseline_res} that the \textbf{IN-OUT} dual embeddings model performed best compared to the others on both datasets, followed closely by the co-purchases baseline. This table proves that the cross embeddings' cosine similarity measure is better able to capture the complementary relationship between products, compared to the conventional technique of using only one embedding \textbf{IN-IN, OUT-OUT}. 

\subsection{Hyperparameter Tuning}

Quite often, \textit{Word2Vec} is used off-the-shelf without optimizing the hyperparameters on the relevant task. While this may be satisfactory when the model is applied to natural language datasets, it is almost always suboptimal when applied to recommender systems datasets. Following previous work on studying the importance of Word2Vec hyperparameters in the recommender systems setting \cite{caselles2018word2vec, 10.1145/3383313.3418486}, we optimize all the relevant parameters of our model, namely the number of negative samples $ns \in (5, 30)$, the negative sampling distribution exponent $\alpha \in (-1, 1)$, the random sub-sampling parameter $t \in (10^{-4}, 10^{-2})$, the initial learning rate $\lambda \in (0.05, 0.15)$ and the embedding dimension $d \in (20, 100)$. We set the number of epochs to a large fixed number because we use early stopping during training. We note that the parameter for the context window size is irrelevant for our setting, since our dataset is in the form of pairs rather than sequences. We optimize the model on a co-purchase prediction task, using a subset (10\%) of the unique co-purchase pairs in the training set. We make sure to remove all of the co-purchase pairs present in the development set from the training set, thus creating a completely disjoint set. The model is essentially tasked with predicting unobserved co-purchase relationships between products. The development set co-purchase pairs are aggregated to have a single target product as the input and the ranked list of frequently co-purchased products as the ground truth. At evaluation time, we query the model with the target item and measure the recall and precision at the top $k$ outputs (nearest neighbors).

Our tuned \textbf{IN-OUT} model on the Overstock dataset has the following hyperparameters: $ns = 20, d = 50, \alpha = -0.1, \lambda = 0.045, t = 0.0001$ and for the Instacart dataset: $ns = 30, d = 100, \alpha = 0.0, \lambda = 0.015, t = 0.001$. Table \ref{tab:hyperparamcompare} shows the relative increase of the evaluation metrics between the optimal and default values. The gains seen in both of the datasets further emphasizes the importance of properly tuning the \textit{Word2Vec} model.

{\renewcommand{\arraystretch}{1.25}
\begin{table}[!ht]
\caption{Relative increase in optimized hyperparameters compared to default values}
\begin{center}
\begin{adjustbox}{width=0.46\textwidth}
        \begin{tabular}{lcccc}
        \toprule
         \multirow{2}{*}{\bf Values} & \multicolumn{2}{c}{\bf Overstock} & \multicolumn{2}{c}{\bf Instacart} \\ \cline{2-5}
         & Precision@20 & Recall@20 & Precision@20 & Recall@20 \\
         \hline
         Default & 0.0236 & 0.1271 & 0.0101 & 0.0173\\
         Optimized & 0.0391 & 0.3229 &0.0437 & 0.0702 \\
         Rel. Increase & 65.6\%& 154\% & 332.6\% & 300\% \\
         \hline
    \end{tabular}
\end{adjustbox}
    \label{tab:hyperparamcompare}
\end{center}
\end{table}}

\subsection{Data Augmentation Results}

In order to better understand the effects of the data and inference augmentation techniques, we split our validation set into two parts. One part only contains target products that appear in the original (unaugmented) training set, this is referred to as the \textit{in-coverage}, the second part is composed of target products that do not appear in the training set, we refer to this portion as the \textit{out-of-coverage} set. While the primary goal of the data augmentation is to provide improvement on the out-of-coverage item prediction, we are also interested in seeing if there are any improvements on the in-coverage set. As baselines, we consider a popularity based recommender (\textit{Top-Sellers}) and a random recommender (\textit{Random}) that simply randomly samples products from the training set for each target item. These heuristics are good choices as baselines since they do not suffer from coverage issues, as they always produce recommendations on any given target product. The models that have +DA or +IA imply that those models had either data augmentation, inference augmentation, or both. We omit evaluation numbers for IA on in-coverage, since IA is only applicable to out-of-coverage predictions. In order to get similarity measurements on product pairs, we train a separate SGNS model on user click sequences. Each sequence consists of clicks accumulated during a 30-minute session. We optimize the hyperparameters of the model on frequently co-clicked product prediction task, where the co-clicks are collected only from search and navigation pages. The reason we chose this evaluation set is to curb some effects of the feedback loop bias introduced by other product recommender systems that are currently in production.

The results in Tables \ref{table:da-eval} and \ref{table:coverage-eval} suggest that the data augmentation has a positive impact on both the in-coverage and the out-of-coverage subsets in terms of predictive relevancy as well as total coverage. Although, the improvement on the out-of-coverage is modest and what's even more surprising, a simple popularity based recommender is seemingly doing as well as our model. 

We hypothesize that part of the issue is because the added synthetic pairs may be capturing novel relationships that are simply not covered by the existing historical data. To gain more insight into the effects of data augmentation, we create an altered version of the validation set where the prediction task is at the taxonomy level rather than at the product level. We think that if the recommended products from the model have relevant taxonomies to the target product's taxonomy, then the recommended products themselves will be relevant as well. Different from the product level evaluation, we measure the metrics at a smaller cutoff (@1 and @3), since any given taxonomy will at most have a handful of relevant taxonomies, as opposed to products that have a much larger relevant candidate set.

\begin{table}[!ht]
\caption{Taxonomy and Product Coverage}
\begin{center}
\begin{tabular}{lccc}
    \toprule
    Model & Taxonomy Coverage & Product Recommendation \\ 
    & & Coverage\\ \midrule
    IN-OUT & 15.40\%  & 66.5\% \\
    IN-OUT+DA & 17.18\%  & 77.12\% \\
    IN-OUT+IA & 16.20\%  & 75.2\% \\
    IN-OUT+DA+IA & \textbf{17.48\%}  & \textbf{81.08\%} \\ \bottomrule
  \end{tabular}
\end{center}
\label{table:coverage-eval}
\end{table}

\begin{table}[!ht]
\caption{Out-of-Coverage Taxonomy Predictions}
\begin{center}
\begin{adjustbox}{width=0.46\textwidth}
\begin{tabular}{lcccc}
    \toprule
    Model & Precision@1 & Recall@1 & Precision@3 & Recall@3 \\ \midrule
    Random & 0.0292   &  0.0100 & 0.0252 & 0.0301\\
    Top-Sellers & 0.0198   &  0.0161 & 0.0221 & 0.0382 \\
    IN-OUT & 0.0000  & 0.0000 & 0.0000 & 0.0000\\
    IN-OUT+DA & 0.0598 &  0.0417 & 0.0326 & 0.0676\\
    IN-OUT+IA & 0.0512   &  0.0392 & 0.0318 & 0.0661\\
    IN-OUT+DA+IA & \textbf{0.0802}  & \textbf{0.0581}  & \textbf{0.0444} & \textbf{0.0919}\\ \bottomrule
  \end{tabular}
\end{adjustbox}
\end{center}
\label{table:taxo-rec}
\end{table}

The results of the taxonomy prediction task shown in Table \ref{table:taxo-rec} suggest that the data augmentation is making a measurable impact on the out-of-coverage set. We see a clear difference between the popularity baseline and our model at the taxonomy level. This points to our models' ability to provide actually relevant recommendations. This is further confirmed by visual validation of the recommendations as well, we refer the reader to Figure \ref{fig:recs_ooc} for example recommendations provided by the model.

\begin{figure}
    \centering
    \includegraphics[scale=0.29]{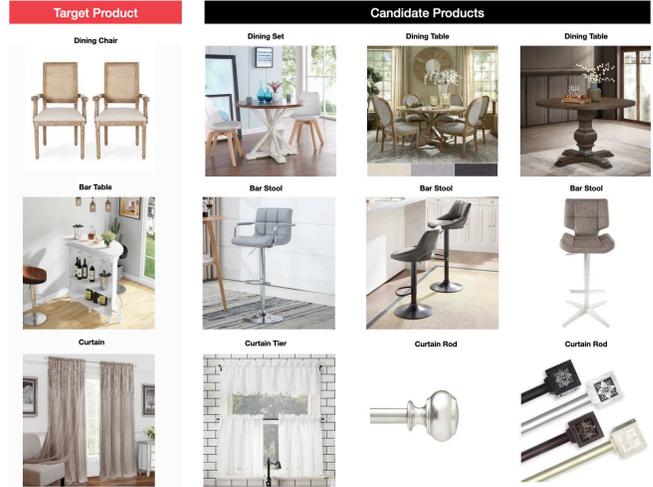}
    \caption{Example candidate sets produced by our model on out-of-coverage target products.}
    \label{fig:recs_ooc}
\end{figure}

\section{Conclusion}
In this paper, we describe an effective method for retrieving complementary products using product embeddings learned with SGNS. Our approach relies on the fact that SGNS learns two separate embeddings for each item. At inference time, we make use of both embeddings to retrieve relevant complementary items. We show that using both of the embeddings rather than the standard approach of only using one significantly improves the results on multiple real world datasets for complementary product recommendations. The advantage of our model is the ease of which it can be implemented and scaled using existing and widely available libraries and software packages. Given enough data, the model can be implemented at a large scale with relatively low latency, especially beneficial for major online e-commerce companies. 

To handle data sparsity issues often exhibited in settings where purchases are used for modeling, we propose a straightforward data augmentation method that has parallels to prior studies in data augmentation for NLP tasks. We rely on similarity measurements learned from other readily available sources of data, such as clicks, to generate novel product pairs. This allows the model to extend its retrieval capability to products that have no or very little purchase data. Our experiments demonstrate that we were able to not only effectively expand product coverage, but also improve relevancy.  We hope that our approach can serve as a strong baseline for future work in the space of complementary product recommendations.

Future studies can explore further development on expanding coverage to even more products without sacrificing, or better yet, improving the performance of the existing covered products. We note that in general, the ability to perform inference on unseen products is an important step in recommendation systems, as newly exposed products will eventually garner more data which will, in time, improve the relevancy of recommended products. In addition, we see a connection between the number of augmented products included in the training set and the concept of exploration vs. exploitation in reinforcement learning. This connection can be investigated further. In particular, an interesting phenomenon which often happens during an exploration phase is that more co-purchases and co-clicks data are obtained for explored items, which can later improve our recommendations. We call this process {\em self-healing} which can be further quantified and measured in the future.

\label{sec:discussions}



\bibliographystyle{abbrv}
\bibliography{reffile}

\end{document}